\documentclass[superscriptaddress,aps,pra,twocolumn,amssymb,amsfonts,showpacs,letterpaper]{revtex4}
\usepackage{amsmath}
\usepackage{color}
\usepackage{graphics,graphicx}
\newcommand{\ket}[1]{\vert #1 \rangle}
\newcommand{\bra}[1]{\langle #1 \vert}

\newcommand{\tr}{\mathop{\mathrm{tr}}\nolimits}
\newcommand{\var}{\textrm{Var}}
\newcommand{\E}{\mathcal{E}}

\newcommand{\Sld}{\Lambda}

\begin{document}
\begin{abstract}
  We introduce an operational interpretation for pure-state global
  multipartite entanglement based on quantum estimation. We show that
  the estimation of the strength of low-noise locally depolarizing
  channels, as quantified by the regularized quantum Fisher
  information, is directly related to the Meyer-Wallach multipartite
  entanglement measure. Using channels that depolarize across
  different partitions, we obtain related multipartite entanglement
  measures. We show that this measure is the sum of expectation values
  of local observables on two copies of the state.
\end{abstract}

\title{An operational interpretation for global multipartite entanglement}
\author{S. Boixo}
\affiliation{Department of Physics and Astronomy, University of New
Mexico, Albuquerque, New Mexico 87131, USA}
\affiliation{Los Alamos National Laboratory, Los Alamos, NM 87545, USA}
\author{A. Monras}
\email{amonras@ifae.es} \affiliation{Grup de Fisica Te\`orica, Universitat Aut\`onoma de Barcelona, Bellaterra, E-08193, Spain}
\maketitle

%\section{Introduction}
Schr\"odinger, inspired by the EPR paper, described an entangled state
in the following terms: ``the best possible knowledge of the whole
does not include the best possible knowledge of its
parts''~\cite{epr_schroedinger1935}. Entanglement has ever since
played a prominent role in foundational studies of quantum mechanics
because of the relation between entanglement and stronger than
classical nonlocal correlations. Nevertheless, only in recent years
has a formal theory of quantum entanglement been developed.  The
reason is that entanglement is a prerequisite for performing
paradigmatic tasks in quantum information, such as teleportation,
dense coding, or pure-state quantum
computation~\cite{nielsen_quantum_2000}. More precisely, the most
common bipartite entanglement measures have an operational
interpretation in terms of a quantification of the available resources
for a specific task ~\cite{horodecki_quantum_2007}. In this Letter a
similar operational interpretation for global multipartite
entanglement is derived.

Entanglement is only a necessary but not sufficient condition for
computational speedups, as shown by the Gottesman-Knill
theorem~\cite{gottesman_stabilizer_1997}.  On the other hand, the
figure of merit of any quantum information processing task can be used
to define an entanglement monotone if the initial state is optimized
with local operations and classical communication (LOCC)
preprocessing~\cite{le}. One particular measure, localizable
entanglement, arose from the study of the suitability of a given state
to perform quantum communication with quantum repeaters. It quantifies
the amount of entanglement attainable between two specific parties
after performing LOCC on the rest. When there is enough symmetry,
localizable entanglement can be analytically computed and indicates
multipartite entanglement~\cite{adesso_equivalence_2005}.

Entanglement also improves the precision of quantum measurements, a
task itself related to quantum computation~\cite{esti}. Instead of
focusing on computational tasks, here we will present an
interpretation of entanglement as a resource for a specific parameter
estimation problem. A good candidate for a parameter invariant under
local unitaries is the strength of a locally depolarizing channel,
\emph{i.e.}, a tensor product of depolarizing channels which mimics
the tensor structure that defines locality for the given parties. It
has been noted that entanglement helps, as expected, in estimating the
parameters of a quantum channel~\cite{channels,
  hotta_n-body-extended_2006}. In the specific case of a two-qubit
locally depolarizing channel, maximally entangled states achieve the
best precision in the estimation for some range of depolarization. On
the other hand, entanglement is not useful for all values of the
depolarization strength, and mixed entangled states tend to perform
worse than separable states~\cite{channels2}. Finally, Fisher
information, a concept central to the quantification of estimation
sensitivity, as we will see, has been found to be proportional to the
logarithmic negativity, in the context of dense coding, for squeezed
states and some particular two-qubit
states~\cite{kitagawa_entanglement_2006}.

The keystone of quantum parameter estimation is the so-called
\emph{quantum Cram\'er-Rao bound} (QCRB)~\cite{crb}. To understand its
meaning we first draw an analogy with the theory of statistical
estimation. A statistical model $M$ is a parametrized family of
probability distributions $M=\{p_\epsilon(x);
\epsilon\in\Theta\}$. Estimators $\hat\epsilon$ are functions of the
outcomes $x$ onto the parameter space $\Theta$. An estimator is
unbiased if $\sum_x \hat\epsilon(x)p_\epsilon(x)=\epsilon$. The
single-parameter Cam\'er-Rao bound \cite{crb} for unbiased estimators
is $\var_\epsilon[\hat\epsilon] I_\epsilon\geq 1$, where $I_\epsilon$
is the \emph{Fisher information} of the model $M$, and $\var_\epsilon[\hat\epsilon]$ is the variance of the estimator,
\begin{equation}
 	\label{eq:Fisher}
	I_\epsilon=\sum_x \left(\frac{\partial \log p_\epsilon(x)}{\partial \epsilon}\right)^2p_\epsilon(x).
\end{equation} 
Note that $I_\epsilon$ provides a measure of distinguishability.

Quantum-mechanically, the statistical model is replaced by the
\emph{quantum model}, \emph{i.e.}, a parameterized family of quantum
states $\mathcal M=\{\rho_\epsilon;\epsilon\in\Theta\}$. While the
classical Fisher information provided by a measurement depends on the
measurement itself, the quantum Fisher information (QFI) $J_\epsilon$
(defined below) does not. The single-parameter QCRB is
$\var_\epsilon[\hat\epsilon] J_\epsilon\geq 1$, and is attainable
asymptotically in the number of measurements. When an estimator
$\hat\epsilon$ attains the QCRB it is said to be \emph{efficient}. The
QFI $J_\epsilon$ provides the quantum model with a geometric structure
of operational significance. It can also be shown that the QFI is
proportional to the Hessian matrix of the quantum fidelity \cite{crb},
$\mathcal F(\rho,\sigma)=\tr[\sqrt{\sqrt{\sigma}\rho\sqrt{\sigma}}]$,
$
	\mathcal F(\rho_\epsilon,\rho_{\epsilon+\phi})=1-\frac{J_\epsilon}{8}\phi^2,
$
hence showing that the quantum Fidelity has a clear interpretation in
terms of distinguishability.

\begin{figure}[t]

%     \subfloat[GHZ States]{
%         \label{fig:ghz}
        \includegraphics[width=0.23\textwidth]{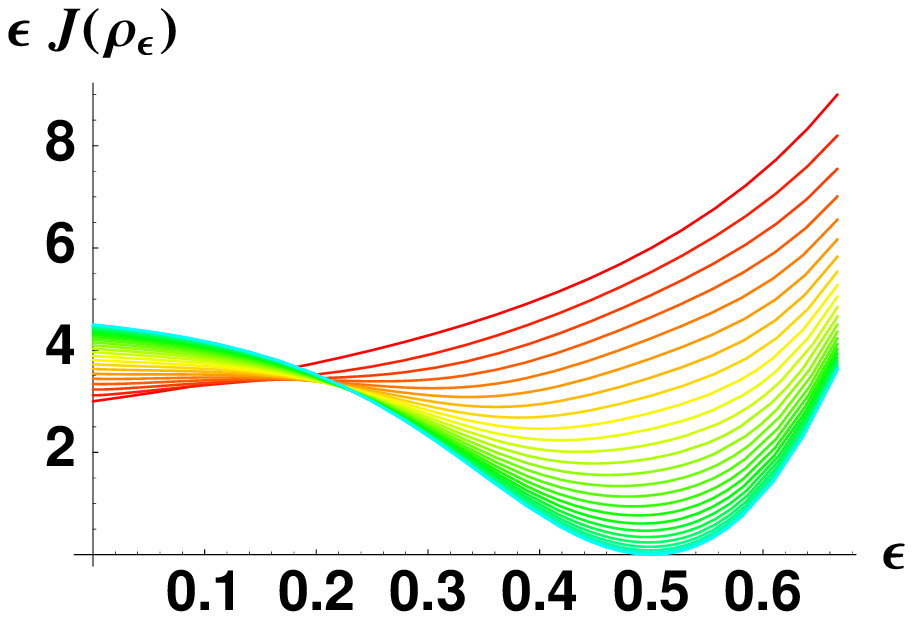}
%     }
%     \subfloat[W States]{
%         \label{fig:w}
        \includegraphics[width=0.23\textwidth]{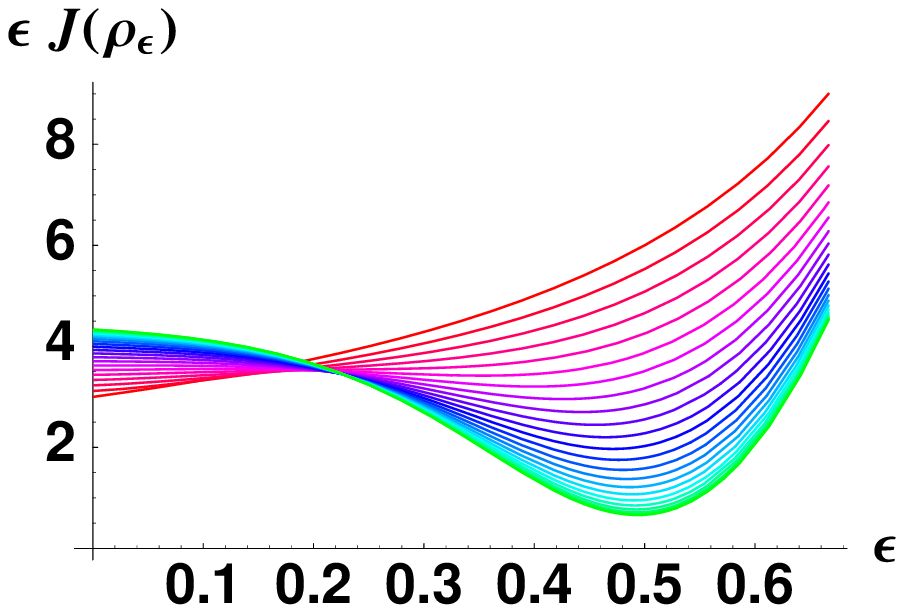}
%    }
    \caption{\footnotesize (Color online). Regularized quantum Fisher
      information as a function of the noise parameter for a locally
      depolarizing channel in three qubits. Left figure shows the
      regularized QFI for GHZ states $\sqrt {\mu_1} \ket {000} + \sqrt
      {1-\mu_1} \ket {111}$ of three qubits. The darkest line (red
      online) represents a separable state and the lightest line
      represents a maximally entangled state. Right figure shows
      the regularized QFI for W states $\mu_1 \ket {100} + \mu_2 \ket
      {010} + \mu_2 \ket {001}$, with $\mu_2 =
      \sqrt{(1-\mu_1^2)/2}$. Intermediate lines correspond to
      intermediate states. \label{figure}}
\end{figure}

Entanglement is a fragile resource under local noise. It is this
feature that gives entangled states their usefulness in loss
estimation. A quantification of this usefulness would entail, in
principle, a means of quantifying the amount of entanglement.  It
turns out that, since entangled states are the ones that decohere
faster, above some threshold value of $\epsilon$ their sensitivity
drops below that of a separable state. This is the transition effect
found by Fujiwara~\cite{channels2} and shown in Fig.~(\ref{figure}),
which plots the regularized QFI, $\epsilon J(\rho_\epsilon)$ as a
function of the strength of the channel $\epsilon$.  A similar effect
is found when analyzing phase estimation in the presence of
decoherence~\cite{deco}.  When one takes a pure-state (separable or
not) through a low-noise channel, the state becomes slightly
mixed. The noise parameter becomes related to the entropy of the state
itself. It turns out that the QFI diverges when one approaches the
boundary of pure states from the set of mixed states, a fact closely
related to the divergence of the Bures metric for pure states, as well
as some surprising results in amplitude damping channel
estimation~\cite{monras_optimal_2007}. This divergence becomes
intuitive when one considers the problem of estimating the parameter
$p$ of a binomial distribution. When such parameter approaches zero,
the variance of the estimation $p(1-p)$ also does at the same rate,
and the QFI also diverges as $1/p$. This is a key signature of
Poissonian statistics.  The discussion above shows that it will be
necessary to appropriately regularize the divergence.

In the case of two qubits there is only one kind of entanglement,
given by the Schmidt coefficients, so before the crossover maximally
entangled states are optimal, and after the crossover separable states
are optimal.  In the case of three qubits the
entanglement of the W states is not the same as the entanglement of
the GHZ states, and the crossover occurs at a slightly different
point, so for some range of $\epsilon$, W states outperform GHZ
states (see Fig.~\ref{figure}). This can be explained by the rate of change of the von Neumann
entropy, initially higher for GHZ states, but later higher for W
states.  Incidentally, GHZ states are also optimal for local phase
estimation without decoherence. On the other hand, the entanglement in
GHZ states is more persistent than in W
states~\cite{simon_robustness_2002}. The von Neumann entropy is flat
for GHZ states around $\epsilon=.5$, which corresponds to the totally
depolarizing channel according to our choice for channel
parametrization -see Eq.~\eqref{eq:onechannel} and \eqref{eq:re}-, and
GHZ states provide no QFI in that region.  That is not the case for W
states. In all cases the rate of change of the von Neumann entropy
diverges when $\epsilon \to 0$, which also explains the divergence of
the QFI. Finally, the QFI increases as the channel becomes invertible
\cite{rungta_universal_2001}, but separable states still outperform
entangled states in that region.

We now show that the QFI for a low-noise locally depolarizing channel
is an entanglement measure. To avoid the difficulties discussed so far
when the state becomes too mixed, we will take the limit as
$\epsilon\to0$, \emph{i.e.}, the channel is close to the identity
channel. We will use the renormalized QFI, $\epsilon
J(\rho_\epsilon)$, to cancel the divergence of the Bures metric in the
transition from pure to mixed states. Due to this divergence, we do
not expect our proposed measure to be a good candidate for detection
of entanglement with mixed states: pure states, even without any
entanglement, will in general perform better than initially
mixed states.

%\section{The measure}
The trace preserving channel that commutes with all unitaries can be
written as~\cite{rungta_universal_2001}
\begin{align}
  \label{eq:onechannel}
  \E_\epsilon(\rho) = (1-d\epsilon) \rho + \epsilon \openone \tr \rho\;,
\end{align}
where $d$ is the dimension of the Hilbert space on which the channel
acts. Up to a choice of $\epsilon$, the channel is unique. The channel
is completely positive in the range $ 0 \le \epsilon \le d
/({d^2-1})$. To first order in $\epsilon$, the channel for $n$ parties
acting on state $\rho$ gives
\begin{align}
  \label{eq:re}
  \rho_\epsilon \equiv \E_\epsilon^{\otimes n}(\rho) = \Big(1-\epsilon\sum_j d_j\Big) \rho + \epsilon \sum _j
  \openone _j \otimes \tr_j\rho\;,
\end{align}
where $\openone_j$ denotes the identity in party $j$ and $\tr_j$
denotes the partial trace with respect to party $j$.

To obtain the QFI, the standard procedure starts by solving for the
symmetric logarithmic derivative $\Sld_\epsilon$, defined as any
Hermitian operator that satisfies the equation
$%\begin{align}
    \label{eq:SLDdef}
  \Sld_\epsilon  \rho_\epsilon + \rho_\epsilon \Sld_\epsilon = 2\partial_\epsilon
  \rho_\epsilon 
$%\end{align}
. The QFI does not depend on the particular choice of $\Sld_\epsilon$, and
is given by (note the clear analogy with Eq. \eqref{eq:Fisher})
\begin{align}
  \label{eq:QFIdef}
  J(\rho_\epsilon) = \tr[ \rho_\epsilon \Sld_\epsilon^2]=\tr[(\partial_\epsilon \rho_\epsilon)\Sld_\epsilon]\;.
\end{align}

The output state can be expanded as $\rho_\epsilon = \rho -
\epsilon\rho'+O(\epsilon^2)$, where
$\rho'=-\left[\partial_\epsilon\rho_\epsilon\right]_{\epsilon=0}$.
Because of the $1/\epsilon$ divergence in the frontier of pure states,
a solution for $\Lambda_\epsilon$, for initial pure states,
is~\cite{hotta_n-body-extended_2006} $\Sld_\epsilon = (\openone -
\rho)/\epsilon -\rho' + O(\epsilon)$.  Substituting in
Eq.~\eqref{eq:QFIdef}, the QFI reads, to leading order,
\begin{equation}
    \label{eq:closedJ}
    J(\rho_\epsilon)= \frac{1}{\epsilon}\tr[\rho\rho']+O(1)\;.
\end{equation}
In the limit $\epsilon\rightarrow0$ the problem becomes classical
since the optimal measurement is independent of $\epsilon$. In fact,
the projection-valued measurement $\{\mathcal O_x\}$ with $\mathcal
O_0=\rho,~\mathcal O_1=1-\rho$, together with
$\hat\epsilon(x)={x}/{\tr[\rho\rho']}$, provide an unbiased and
\emph{efficient} estimator to leading order near $\epsilon=0$, with
$\var_\epsilon[\hat \epsilon] = \epsilon \,( {\tr[\rho \rho'] \nu})^{-1} +
O(\epsilon^2)$, where $\nu$ is the number of samples measured.

We define the entanglement measure as
\begin{align}
  \label{eq:em}
  E(\rho) \equiv K+\lim_{\epsilon \to 0} \epsilon J
  (\rho_\epsilon)=K+\tr[\rho\rho']\;,
\end{align}
where $K=\sum_j(1-d_j)$ is a constant, depending only on the
dimensions of the parties $\{d_1,\ldots,d_n\}$, to ensure that for
separable states $E(\rho_{\mathrm{sep}})=0$. Another interpretation of
this measure can be given by rewriting Eq.~\eqref{eq:em} as $
E(\rho)=K-\lim_{\epsilon\to0}\partial_\epsilon\mathcal
F(\rho,\rho_\epsilon)^2$, where $\mathcal
F(\rho,\sigma)=\tr[\sqrt{\sqrt{\sigma}\rho\sqrt{\sigma}}]$ is the
fidelity. This confirms that the entanglement measure
corresponds to the rate at which the state $\rho_\epsilon$ moves away
from the initial state under the action of a low-noise locally
depolarizing channel. Usually the QFI will correspond to the second
derivative of the fidelity, and the fidelity would have a local
maximum for $\epsilon=0$. In this case, though, the channel is
unphysical for $\epsilon<0$, and the first derivative of the fidelity
at $\epsilon = 0$ does not vanish. This is captured by the divergence
of the QFI.

To proceed, we get an expression for $\rho'$ from Eq.~(\ref{eq:re}), $
\rho' = \sum_j\left(d_j\rho-\openone_j\otimes \rho_j\right)$, where
$\rho_j=\tr_j\rho$.  Plugging back into the definition of the
entanglement measure Eq.~(\ref{eq:em}) we obtain
\begin{align}
  E(\rho) = \sum_j\big(1 - \tr\left[\rho \left(\openone _j \otimes
  \rho_j\right)\right]\big) = \sum_j(1 - \tr[\rho_j^2])\;.
\end{align}
The final entanglement measure is just the sum of local linear
entropies. Up to normalization, this is the Meyer-Wallach multipartite
entanglement measure, itself a special case of Generalized
Entanglement~\cite{ge}. We have shown that the precision of the
estimation of the strength of a low-noise locally depolarizing channel
is given by the global multipartite entanglement of the initial
state. Notice, though, that this procedure does not detect {\em
  genuine} multipartite entanglement~\cite{coffman00}.

Different entanglement measures can be derived using channels with
different tensor structures. For a selection of parties
$\alpha=\{\alpha_1, \dots, \alpha_k\}$, consider the depolarizing
channel for those parties $ \E^\alpha_\epsilon = (1-\epsilon
d_\alpha)\rho + \epsilon \openone_\alpha \otimes \rho_\alpha $.  The
corresponding QFI is, up to additive constants, $J_\alpha(\rho)
\approx 1 -\tr[\rho_\alpha^2]$. When composing channels which
depolarize with respect to different partitions, the channels commute
to first order in $\epsilon$, so the order of the composition is not
important, and the QFI is, up to constants, the sum of the
corresponding local linear entropies.  For instance, the composition
of the depolarizing channels for all partitions $
\E^{\mathcal{N}}_\epsilon = \E^{\alpha_1}_\epsilon \circ \ldots \circ
\E^{\alpha_\mathcal{N}}_\epsilon$, gives an entanglement measure
\begin{align}
      E_p(\rho) = K_p + \lim_{\epsilon \to 0} \epsilon J_p
      (\rho_\epsilon)=\sum_\alpha(1-\tr[ \rho_\alpha^2])\;,
\end{align}
where $\alpha$ runs over all partitions. This measure is proportional
to a generalization of the Meyer-Wallach entanglement measure
~\cite{mintert_concurrence_2005}. Similar measures have been used in
the context of quantum phase
transitions~\cite{de_oliveira_multipartite_2006}. Here we will not
consider this extensions any further, but the following analysis
applies trivially.

Pure-state entanglement measures can be extended
to mixed states by the convex roof,
\begin{align}\label{eq:conroof}
  E(\rho) \equiv \min_{\{p_j, \ket {\Psi_j} \} } \Big\{ \sum_j p_j
    E(\ket{\Psi_j}) \Big\}\;,
\end{align}
where $\rho = \sum_j p_j \ket {\Psi_j} \bra {\Psi_j}$. The convex roof
extension can be understood as the solution to a zero-sum two-player
game: system and ``environment''. Let the parties share a mixed state
of the system, $\rho$, and Eve hold a purification of $\rho$. The
parties want to optimize their estimation of the channel while Eve
aims at minimizing the amount of information. Eve is allowed to
perform any rank $1$ measurement on her purification but has to
communicate the classical outcome to the parties. Let $\ket{\Psi_j}$
be the state that the parties are left with after Eve's measurement,
with probability $p_j$. The expected QFI obtained by the parties is
$\sum_j p_jJ(\E_\epsilon(\ket{\Psi_j}))$. On the other hand, by virtue
of the HJW theorem~\cite{hughston_complete_1993}, Eve can prepare any
ensemble $\{p_j,\ket{\Psi_j}\}$ such that $\rho=\sum_j p_j\ket{\Psi_j}
\bra{\Psi_j}$. The minimization performed by Eve will result in an
expected QFI which immediately translates into Eq.~\eqref{eq:conroof}.

%\section{Properties}
We proceed to note some properties of this entanglement
measure. Invariance under local unitaries follows from the symmetry of
the channel. It is also invariant when adding a pure local
ancilla. Strong monotonicity means that $ E(\rho) \ge \sum_j p_j
E(\sigma_j) $, where $\{p_j, \sigma_j\}$ is any ensemble obtained from
$\rho$ with LOCC~\cite{horodecki_quantum_2007}. This prevents
$E(\rho)$ from increasing with LOCC. It is also desirable that the
entanglement measure does not increase when information is lost, that
is, for any ensemble $\sum_j p_j \tau_j =\rho$, $ E(\rho) \le \sum_j
p_j E(\tau_j)$.  For multipartite convex roof extensions derived from
bipartite pure-state entanglement measures it is sufficient to verify
that the local bipartite function is concave in order to prove the
above properties. The concavity of the local linear entropy has
already been shown~\cite{rungta_universal_2001}.

For bipartite states, the entanglement measure given by
Eq.~\eqref{eq:conroof} is known as the
\emph{tangle}~\cite{rungta_universal_2001}. The tangle is the convex
roof of the square of a generalization of the
concurrence~\cite{wootters_entanglement_1998}, derived through the
universal inverter $\mathcal{S} \propto \mathcal P_{\openone} -
\mathcal{I}$, where $\mathcal P_{\openone}$ is proportional to the
projection superoperator onto the identity operator, and $\mathcal{I}$
is the identity superoperator. For pure states the tangle is, up to
additive constants, $ \tr[\rho \mathcal{S}^{\otimes 2}(\rho)] \approx
- \tr[\rho (\mathcal P_{\openone} \otimes \mathcal{I} + \mathcal{I}
\otimes \mathcal P_{\openone} )(\rho)]$.  Now, for the depolarizing
channel, $\partial _\epsilon \E_\epsilon^{\otimes 2}\big|_{\epsilon
  \to 0} = \mathcal P_{\openone} \otimes \mathcal{I} + \mathcal{I}
\otimes \mathcal P_{\openone}$, and $ E(\rho) = K - \tr[
\rho \partial_\epsilon \mathcal{E_\epsilon}^{\otimes
  2}(\rho)\big|_{\epsilon \to 0}]$, where $K$ fixes the relevant
constants. This shows the relation between the QFI of the locally
depolarizing channel and the universal inverter.

%\section{Measuring the Entanglement}
We now introduce an observable whose expectation value gives, up to
normalization, the quantity $E(\rho)$. Let us assume that the parties
have access to many copies of the same pure state.  Further, we assume
that they can perform repeated collective measurements on pairs of
states. Because $E(\rho)$ is a quadratic function, this will be
enough~\cite{mintert_concurrence_2005, measuring}. In particular,
generalizing the expression for the bipartite tangle
from~\cite{walborn_experimental_2006}, we can write
$%\begin{align}
  1-\tr[\rho_j^2]   =  2 \bra \Psi \bra \Psi P_j^-  \ket \Psi \ket \Psi\;,
$%\end{align}
where $P^-_{j}$ is the projector onto the antisymmetric subspace of the
$j$th local Hilbert space of the two copies. The sum of linear
entropies is then
$%\begin{align}
    E(\rho) = \sum_j 2 \bra \Psi \bra \Psi P^-_{j} \ket \Psi \ket \Psi\;,
$%\end{align}
showing that $E(\rho)$ is a sum of expectation values of local
observables, where locality refers to the parties (not the copies).
This measure has been implemented experimentally for two-qubit
states~\cite{walborn_experimental_2006}.

%\section{Conclusion}
In conclusion, while the bipartite entanglement of a state has a quantitative
operational interpretation as the number of qubits that can be
teleported using that state, a similarly clear interpretation has been
lacking for multipartite entanglement. In this Letter we have proposed a quantitative operational
interpretation for global multipartite entanglement as the enhancement on the
estimation of the strength of a low-noise locally depolarizing channel. The
estimation is, by construction, invariant under local unitaries, and
embodies the appropriate tensor structure. 
The variance of the estimation is
related to the rate of change of the von Neumann entropy, and,
therefore, to decoherence. Technical considerations show that the
right interpretation is derived from the regularized quantum Fisher information in the low-noise 
limit. This gives an entanglement monotone proportional to the
Meyer-Wallach entanglement measure. Low-noise depolarizing channels with
different tensor structures give related entanglement measures.
The Meyer-Wallach entanglement measure reduces to the sum of the
averages of local projectors, and might be implementable with current
technology, as has been done already for the bipartite case.

%\section{Acknowledgments}

We wish to acknowledge G. Adesso, E. Bagan, H. Barnum, C. M. Caves,
A. Datta, M.  Elliott, S. Flammia and L. Viola for useful
suggestions. This work was partially carried out under the auspices of
the National Nuclear Security Administration of the US Department of
Energy at Los Alamos National Laboratory under Contract
No. DE-AC52-06NA25396 and partially supported by ONR Grant
No. N00014-07-1-0304, MCyT Project No. FIS2005-01369, and
Consolider-Ingenio 2010, project ``QOIT''.

\end{document}